\documentclass[prb, twocolumn,amsmath,amssymb]{revtex4}
\usepackage{graphicx, color}% Include figure files
\usepackage{dcolumn}% Align table columns on decimal point
\usepackage{bm}% bold math
\begin{document}
\title{A New Platform for Engineering Topological Superconductors: Superlattices on Rashba Superconductors} 
\author{Yao Lu$^*$, Wen-Yu He}\thanks{Equal contribution}
\author{Dong-Hui Xu, Nian Lin}
\author{ K. T. Law}\thanks{Correspondence address : phlaw@ust.hk}
\affiliation{Department of Physics, Hong Kong University of Science and Technology, Clear Water Bay, Hong Kong, China}
\date{\today}
\begin{abstract}
{\bf The search for topological superconductors which support Majorana fermion excitations has been an important topic in condensed matter physics. In this work, we propose a new experimental scheme for engineering topological superconductors. In this scheme, by manipulating the superlattice structure of organic molecules placed on top of a superconductor with Rashba spin-orbit coupling, topological superconducting phases can be achieved without fine-tuning the chemical potential. Moreover, superconductors with different Chern numbers can be obtained by changing the superlattice structure of the organic molecules.}
\end{abstract}

\maketitle
\section{\bf Introduction}
Topological superconductors are new states of matter which are gapped in the bulk and support gapless Majorana fermion excitations at the boundary [\onlinecite{Wilczek, KH, Moore,QZ, Alicea1, Beenakker, Franz, ST}]. It has been pointed out that Majorana fermions are self-Hermitian particles and obey non-Abelian statistics with potential applications in fault-tolerant quantum computations [\onlinecite{ Kitaev1, Kitaev2, NSSFS}] and in spintronics [\onlinecite{ James1, James2}]. Due to the exotic properties of Majorana fermions, the search for topological superconductors has been an important topic in recent years.

It was first pointed out that Majorana fermions can exist in chiral p-wave superconductors where spin-polarized electrons are paired to form Cooper pairs [\onlinecite{Read, Kitaev1}]. It was shown by Kitaev that the essence of creating chiral p-wave superconductor is to pair electrons from an odd number of partially occupied subbands at the Fermi energy [\onlinecite{Kitaev1}]. However, intrinsic chiral p-wave superconductors are yet to be identified. 

After the discovery of topological insulators, Fu and Kane proposed that conventional s-wave superconductors in proximity to a topological insulator surface can be used to engineer p-wave topological superconductors [\onlinecite{Fu}]. Their proposal was based on the fact that:  1) an odd number of partially occupied surface bands exist on the surface of topological insulators, and 2) the surface electrons are helical, in the sense that electrons with opposite momentum have opposite spin due to spin orbit couplings (SOC). As a result, superconducting pairings can be effectively induced by an s-wave superconductor through proximity effect. The resulting s-wave superconductor can be mapped to an effective p-wave topological superconductor. This proposal has been pursued by several experimental groups [\onlinecite{Xue,Yao, Burch}] but it is a challenge to distinguish the Majorana fermions from other low energy fermionic excitations in the system.

Another promising way to create a chiral p-wave superconductor is to apply a Zeeman field to a semiconductor with Rashba spin-orbit coupling (SOC) and induce s-wave pairing through proximity effects [\onlinecite{STF,SLTD, LSD, Alicea2, ORV}]. First, the Zeeman field opens an energy gap at the $\Gamma$ point such that there is only one partially filled subband at Fermi energy if the Fermi energy is inside the Zeeman gap as depicted in Fig.1a. Second, the partially occupied subbands have helical properties due to Rashba SOC. As a result, pairing of the helical electrons can also be induced by an s-wave superconductor and result in an effective spinless p-wave superconductor. Experiments following this scheme have been performed but the results are not yet conclusive [\onlinecite{ Kouwenhoven, Deng, Heiblum, Aguado}]. One of the most challenging parts of the experiment is to fine tune the chemical potential such that it falls within the Zeeman gap at the $\Gamma$ point, where the size of the Zeeman gap is in the order of meV. 

\begin{figure}
\begin{center}
\includegraphics[width=3.2in]{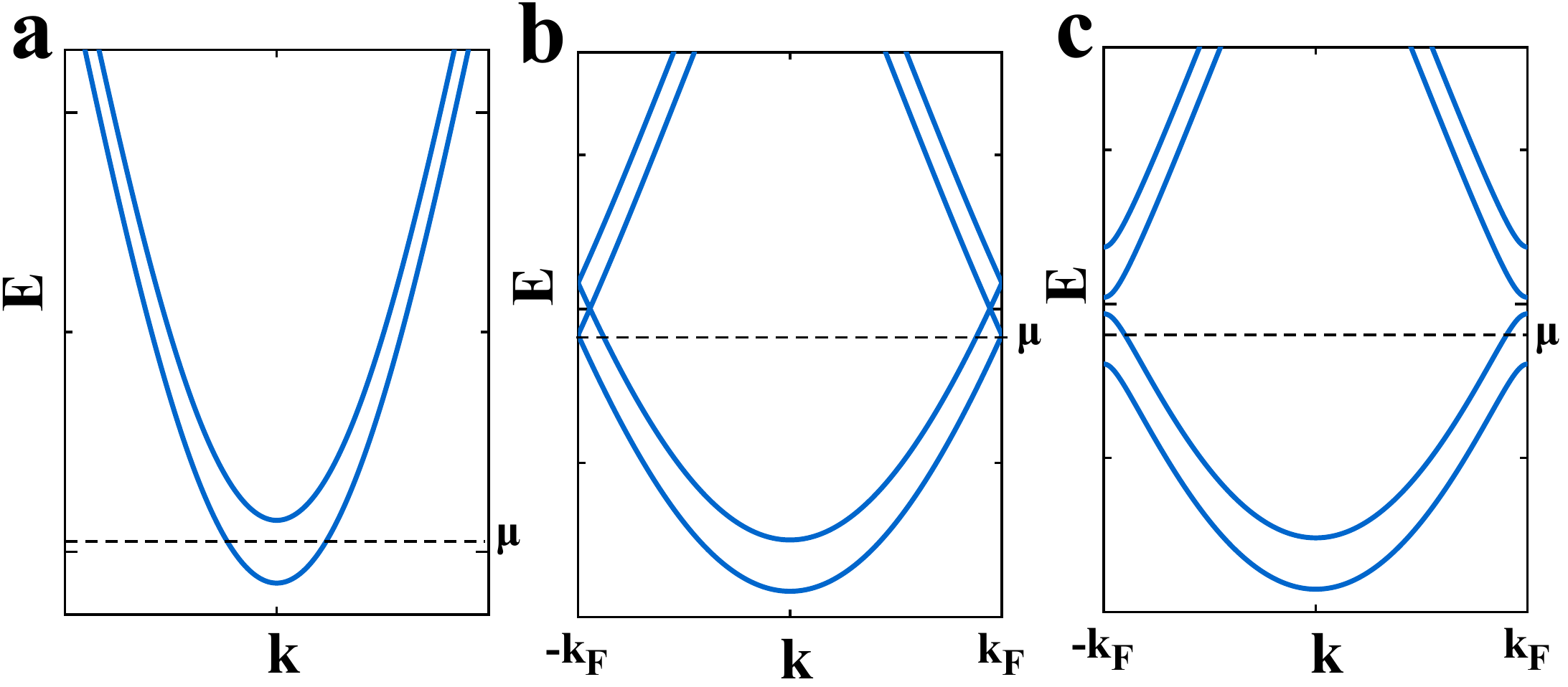} 
\caption{Schematic 1D band structures (a) In the presence of Rashba SOC and a Zeeman field. When the chemical potential $\mu$ is in the Zeeman gap, the outer subband is partially filled. (b) In the presence of a Zeeman field, small Rashba SOC and an infinitesimal superlattice potential. The superlattice potential with period $ \pi/k_F$ reduces the Brillouin zone to $[-k_F, k_F]$. (c) With finite superlattice potential, an energy gap opens at the Brillouin zone boundary and results in a single partially occupied band at the Fermi energy. Inducing superconductivity at the Fermi energy will result in an effective p-wave superconductor. }
\label{Fig1}
\end{center}
\end{figure}

In this work, we propose that putting superlattices on top of a system with Rashba split bands can be used to realise effective p-wave superconductors \emph{without} fine-tuning the chemical potential. To illustrate the ideas, we first consider a one dimensional system with Rashba split bands and where the Fermi energy is far away from the band bottom as illustrated in Fig.1b. We demonstrate how a superlattice potential can open an energy gap for one of the spin subbands at the Brillouin zone boundary such that there is only one partially occupied helical subband at the Fermi energy, as depicted in Fig.1c. Inducing s-wave superconductivity on the system will result in a topological superconductor.

Second, we consider a two dimensional system with Rashba split bands. We show that superlattices, together with time-reversal symmetry breaking, change the Chern number of certain subbands of the system and make the system topological. Interestingly, we show that states with different Chern numbers can be engineered using different superlattice structures and different lattice spacing. Square, rectangular and triangular superlattices are studied in detail as depicted in Fig.3 and Fig.4. Importantly, recent experiments have demonstrated that organic molecules can induce superlattice potentials on metal surfaces to create different band structures such as Dirac band structures [\onlinecite{Manoharan, Nian}]. We believe that inducing superlattice potential and s-wave superconductor on Au or Bi surfaces which possess Rashba surface bands, can be used to create topological superconductors without fine-tuning the chemical potential.

\section{\bf 1D realization}
\emph{Schematic illustration}--- To illustrate the importance of band folding introduced by superlattices, we first consider a wire with Rashba spin-orbit coupling. In the basis of $(\psi_{k \uparrow},\psi_{k \downarrow})$, the Hamiltonian of the wire in the presence of a magnetic field and s-wave pairing can be written as [\onlinecite{STF,SLTD, LSD, Alicea2, ORV}]:
\begin{equation}
H_{1D}=(\frac{k^2}{2m}-\mu)\tau_z+V_x\tau_z\sigma_x+\alpha_Rk\tau_z\sigma_y+\Delta\tau_y\sigma_y
\end{equation}
Here, $m$ is the effective mass of the electrons, $\alpha_R$ is the Rashba SOC strength, $V_x$ is the Zeeman energy from an external magnetic field along the wire. As pointed out previously, the wire becomes a topological superconductor when the chemical potential is inside the Zeeman gap such that $|\mu| < \sqrt{V_x^2-\Delta^2} $, as depicted in Fig.1a. In this case, the chemical potential intercepts a single non-degenerate band and the system can be mapped to a spinless p-wave superconductor [\onlinecite{STF,SLTD, LSD, Alicea2, ORV}]. However, it is very difficult to gate the chemical potential to the Zeeman gap experimentally.

In this section, we point out that the system can become a topological superconductor without fine-tuning the chemical potential. The key is to introduce a superlattice potential $u(x)$ with period approximately equal to $2\pi/k_{F}$ where the $k_F$ is the Fermi wave vector. 

To be specific, we first discuss the case with weak Rashba SOC such that $\alpha_R k_F \ll V_x \ll t$. These parameters are applicable to realistic semiconducting wires used in recent experiments [\onlinecite{Kouwenhoven, Heiblum, Deng, Aguado, Jie}]. In this case, we have two subbands separated by the Zeeman energy. When the superlattice potential is infinitesimal, the size of the Brillouin zone is reduced such that $\pm k_F$ become the zone boundary points as illustrated in Fig.1b. However, when the superlattice potential is finite, it couples states with the same spin at $k$ and $k \pm G$, where $G=2k_F$ and $u_G$ is the corresponding Fourier component of $u(x)$. This can be easily seen in the second quantized form whereby the superlattice potential term can be written as $u_{G}\psi^{\dagger}_{k \sigma}\psi_{k \pm G \sigma}$. 

This coupling will result in an energy gap at the reduced Brillouin zone boundary as depicted in Fig.1c. Denote $|\psi (k) \rangle$ as an eigenstate of $H_{1D}(k)$ with $\Delta=0$, the energy gap at the reduced zone boundary is  $ \langle \psi(k_F)| U_{G} |\psi (-k_F) \rangle =  \frac{U_G V_x(V_x+\sqrt{V_x^2+\alpha_R^2G^2/4})}{V_x^2+\alpha_R^2G^2/4+V_x\sqrt{V_x^2+\alpha_R^2G^2/4}}$. From Fig.1c, we see that if the Fermi energy is within certain energy gaps of the Brillouin zone boundary, there is only one partially filled subband at the Fermi energy. Further introducing an s-wave pairing into the system will result in a topological superconductor. It is important to note that the key to reach the topological regime is to introduce a superlattice potential with appropriate lattice constant.

\begin{figure}
\begin{center}
\includegraphics[width=3.2in]{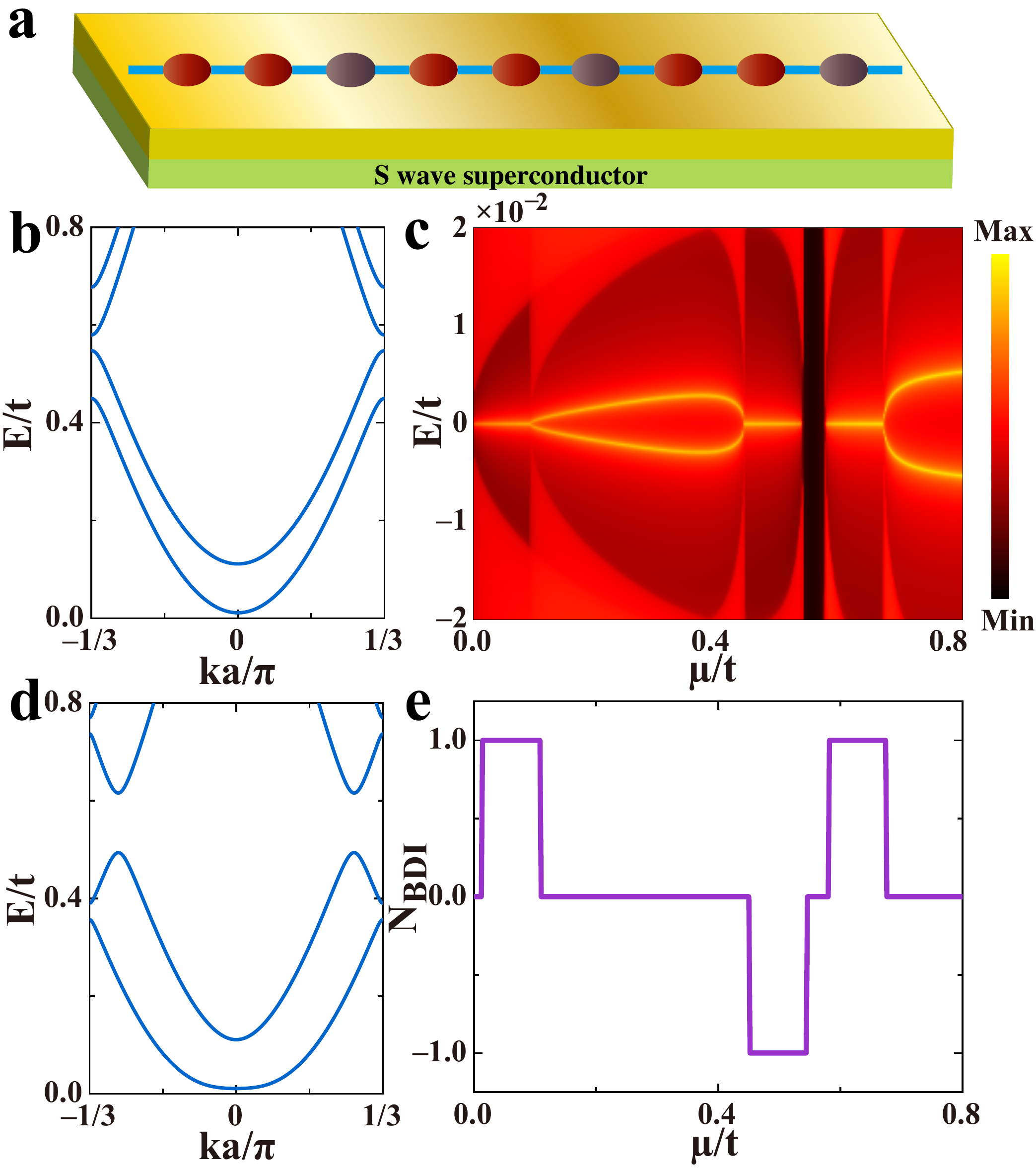}
\caption{(a) The schematic illustration of a 1D wire with the potential of every third site is modified. (b) The spectral function of the last site of $H_{1DTB}$ with $\Delta=0$, $t=1$, $\alpha_R=0.01t$ and $V_x=0.05t$. (c) The spectral function $\mathbf{A}(\omega)$ with parameters as in (b) except $\Delta=0.015t$. (d) The normal state band structure with $\alpha_R=0.2t$ and $V_x=0.05t$. (e) The topological invariant $N_{BDI}$ of $H_{1DTB}$. The topological regimes in (c) and (e) coincide.}
\label{Fig2}
\end{center}
\end{figure}

\emph{Tight-binding calculations}--- To verify the above claims, we construct a tight-binding model to describe $H_{1D}$ and the superlattice potential. To be specific, the superlattice potential is simulated by changing the on-site energy of one site out of every three sites of the original wire as illustrated in Fig.2a. The model can be written as:
\begin{eqnarray}
 H_{1DTB} & = & \sum\nolimits_{i, \alpha } { - t( \psi _{i + 1,\alpha }^\dag  \psi _{i,\alpha }  + h.c.) - \mu \psi _{i, \alpha }^\dag  \psi _{i,\alpha } }  \nonumber \\
&+& \sum\nolimits_{i, \alpha ,\beta } { - i{\alpha _R} \psi _{i +1,\alpha }^\dag  (\sigma_{y})_{\alpha \beta }   \psi _{i,\beta } } \nonumber \\
 & + & \sum\nolimits_{i,\alpha ,\beta } { \psi _{i, \alpha }^\dag  (V_x \sigma_x)_{\alpha \beta} \psi _{i,\beta } }   \nonumber \\
& + &  \sum\nolimits_{i,\alpha} \Delta \psi _{i,\alpha }^{\dagger} \psi _{i,-\alpha }^{\dagger} + u\psi_{3i,\alpha}^{\dagger}\psi_{3i,\alpha}+h.c.  
\end{eqnarray}
Here, $i$ denotes the lattice sites and the spin indices are denoted by $\alpha$ and $\beta$, $t$ and $\mu$ are the hopping amplitude and the chemical potential respectively, $\alpha_{R}$ is the Rashba coupling strength, $V_{x}$ is the Zeeman energy, $\Delta$ is the s-wave pairing amplitude, and $u$ is the on-site potential which is non-zero for every three sites.

As a result of band folding, the new Brillouin zone boundary is located at $\pm \pi/3a$ where $a$ is the lattice spacing of the tight-binding model. The energy spectrum of the model in the absence of superconducting pairing is shown in Fig.2b. As expected, energy gaps open at the Brillouin zone boundary as well as the $\Gamma$ point. The system is expected to be a topological superconductor when the chemical potential is within the regime where an odd number of subbands are partially occupied.

To verify the existence of the Majorana end states of $H_{1DTB}$, we plot the spectral function $\mathbf{A}(\omega)=\text{Tr}[Im \mathbf{G}(\mathbf{R}, \mathbf{R}, \omega)]$ of the last site $\mathbf{R}$ of a semi-infinite wire as a function of chemical potential and energy in Fig.2c. Here, $\mathbf{G}$ is the lattice Green function of $H_{1DTB}$ [\onlinecite{Potter}]. From the spectral function, it is evident that zero energy modes, which correspond to Majorana end states of the wire, appear at the regime where an odd number of subbands are partially occupied. The topological regime at low chemical potential is due to the opening of the Zeeman gap near the band bottom which would appear even in the absence of the superlattice potential. On the other hand, the topological regime created by the opening of an energy gap at the Brillouin zone boundary at the Fermi energy is due to the superlattice potential.

\emph{Topological invariant}--- To further study the topological properties of the system, we write $H_{1DTB}$ in the momentum basis. Due to the superlattice potential, there are three sublattices in each unit cell. The basis can be written as $(\Phi_{k,\uparrow}, \Phi_{k,\downarrow}, \Phi_{-k,\uparrow}^{\dagger}, \Phi_{-k,\downarrow}^{\dagger})$, where $\Phi_{k,\sigma} = [\psi_{k,\alpha}(1),\psi_{k,\alpha}(2),\psi_{k,\alpha}(3)]$ and $\psi_{k,\sigma}(j)$ denotes an electron annihilation operator of the $j$ sublattice. In this basis, $H_{1DTB}$ is in the BDI class due to the presence of the particle-hole symmetry $P=\tau_xK$ and the chiral symmetry $C=\tau_x$, where $K$ is the complex conjugate operator and $\tau_x$ operates on the particle-hole sector. As a result, we can classify $H_{1DTB}$ by a topological invariant $N_{BDI}$ where $N_{BDI}$ is defined in the Method section. The topological invariant $N_{BDI}$ as a function of the chemical potential is shown in Fig.2e. By comparing the topological regime in Fig.2e with the gap opening regime at Brillouin zone boundary in Fig.2b, it is evident that the system is topologically non-trivial when an odd number of subbands are partially occupied.

We note that introducing a superlattice potential can also create topological regimes at the Fermi energy when Rashba energy is strong. For example, the normal state spectrum for $H_{1DTB}$ with strong Rashba SOC such that $\alpha_R k_F \gg V_x$ is illustrated in Fig.2d. In this case, due to the large Fermi momentum difference between the two subbands, the lower subband with larger $k_F$ is gapped out at the Brillouin zone boundary by the superlattice potential. On the other hand, the upper subband remains gapless. The system is topologically non-trivial when an odd number of subbands in the normal state are partially occupied and when s-wave superconductivity is induced.

\section{\bf superlattice in 2D systems}

The above section illustrated how 1D superlattices with suitable lattice spacing can be used to engineer a 1D topological superconductor. In this section, we demonstrate that 2D superlattices on metal surface states can be used to engineer topological superconductors with different Chern numbers without fine-tuning the chemical potential.

Our proposal is inspired by a recent experiment in which coronene molecules were deposited on top of copper [111] surfaces [\onlinecite{Nian}]. The superlattice potential induced by the molecules creates a honeycomb lattice structure on the surface and results in Dirac spectrum of the surface states. The positions of the coronene molecules on the copper surface can be manipulated by the STM tips so that the lattice spacing and structures of the superlattice potential are highly tuneable. Several types of artificially made defect states and zero energy states associated with the zig-zag edge of the sample, were observed, demonstrating the great tuneability of the setup. With this recent advancement in experimental technique, we believe that desirable superlattice potential can be engineered on metal surfaces with strong Rashba surface states as well.

\emph{Square lattice}--- To be specific, we consider Au(111) surfaces and the Au thin layer is placed on top of a superconducting substrate as illustrated in Fig.3a. It is shown both theoretically and through ARPES experiments that the surface states have large Rashba splitting of about 100meV [\onlinecite{Au}]. It was proposed that Au(111) surface states, when combined with Zeeman field and superconductivity, can be used to generate topological superconductors. However, the chemical potential has to be tuned to the Zeeman gap near the Rashba band bottom [\onlinecite{Andrew}]. Since the Fermi energy of Au is about 0.4eV above the Rashba band bottom [\onlinecite{Au}], it is very difficult to gate the chemical potential to the Zeeman gap. As we show below, a suitable superlattice potential and a Zeeman field can make the Au thin film a topological superconductor.

\begin{figure}
\begin{center}
\includegraphics[width=3.2in]{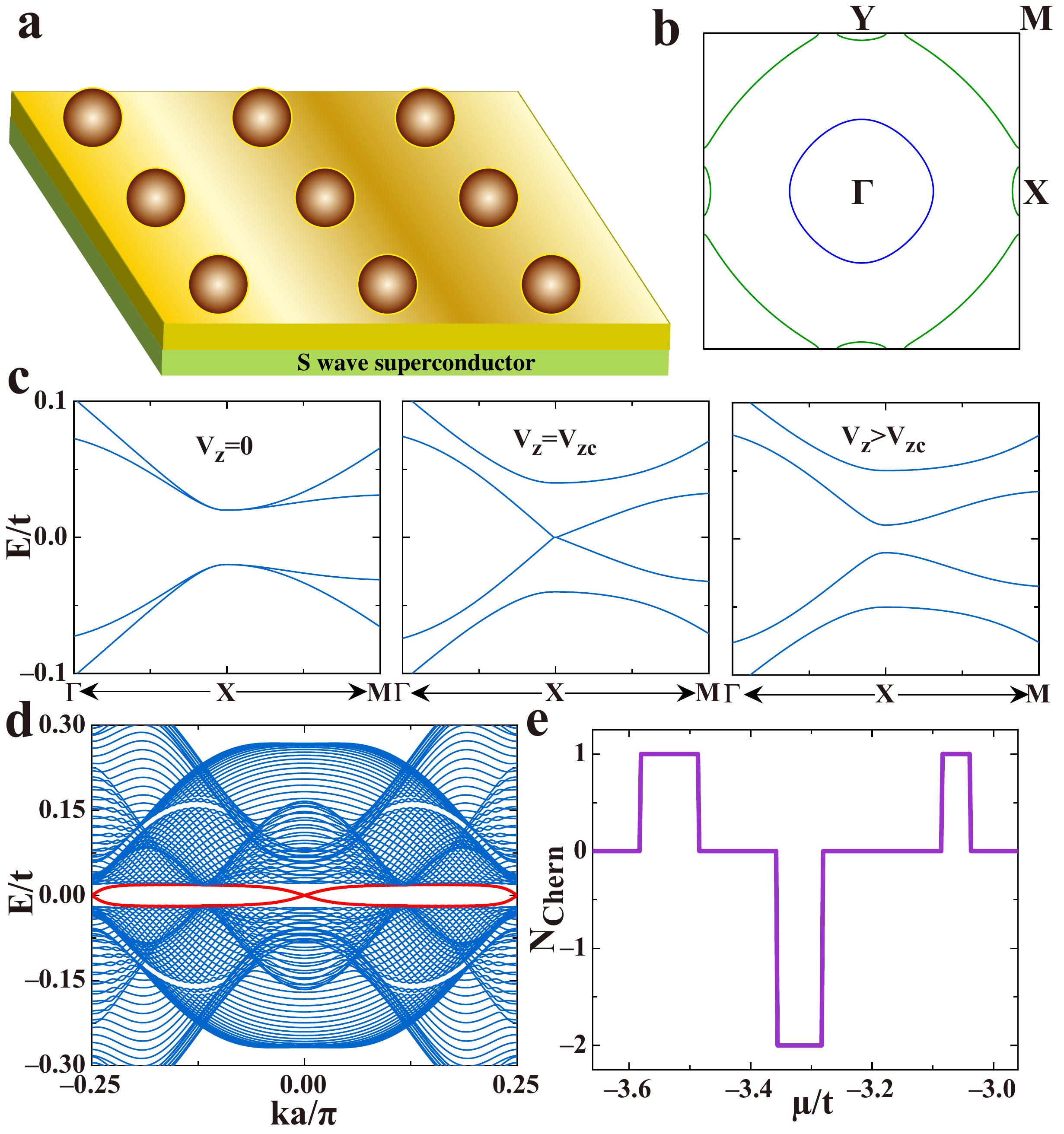}
\caption{ (a) Organic molecules on the top of Au (111) surface where Au is grown on top of an s-wave superconductor. (b) The outer (green) and inner Fermi (blue) circles in the superlattice Brillouin zone. The square indicates in reduced Brillouin zone due to superlattice potential.  (c) The band closing and reopening at X point, as a function of $V_z$. $V_{zc}$ denotes the critical value of $V_z$ at which the band gap closes. (d) The energy spectrum of Au surface with square superlattice potential. The tight-binding model  $H_{2DTB}$ used is described in the Method section. The superlattice is chosen such that the Chern number is -2. It is evident that there are two crossings at zero energy corresponding to the two zero energy Majorana modes at each edge. (e) The Chern number of $H_{2DTB}$ as a function of chemical potential for a fixed superlattice configuration. }
\label{Fig3}
\end{center}
\end{figure}

We first place organic molecules on top of the Au(111) surface to induce a superlattice potential as illustrated in Fig.3a. If the superlattice potential has a square structure with lattice spacing $\pi/k_F$, where $k_F$ is the Fermi wave vector of the Rashba band with larger Fermi momentum, the X and Y points $(\pm k_F, 0)$, $( 0, \pm k_F)$ becomes the Brillouin zone boundary points as depicted in Fig.3b. From ARPES experiments, $k_F$ is about $0.2/\mathring{A}$ [\onlinecite{Au}] and the superlattice spacing should be about 1.6nm. This superlattice spacing falls within the experimentally accessible regime. 

Now, if an s-wave pairing is induced on Au(111) surface, a pairing gap opens at the Fermi energy as illustrated in Fig.3c. The energy spectrum in Fig.3c is based on a tight-binding model introduced in the Method section. Due to the pairing gap, one may classify this two dimensional system by Chern numbers. However, due to time-reversal symmetry, the total Chern number of the occupied bands have to be zero. Indeed, from the tight-binding model calculations, one can show that the outer and inner occupied bands have Chern number 1 and -1 respectively.

\begin{figure}
\begin{center}
\includegraphics[width=3.2in]{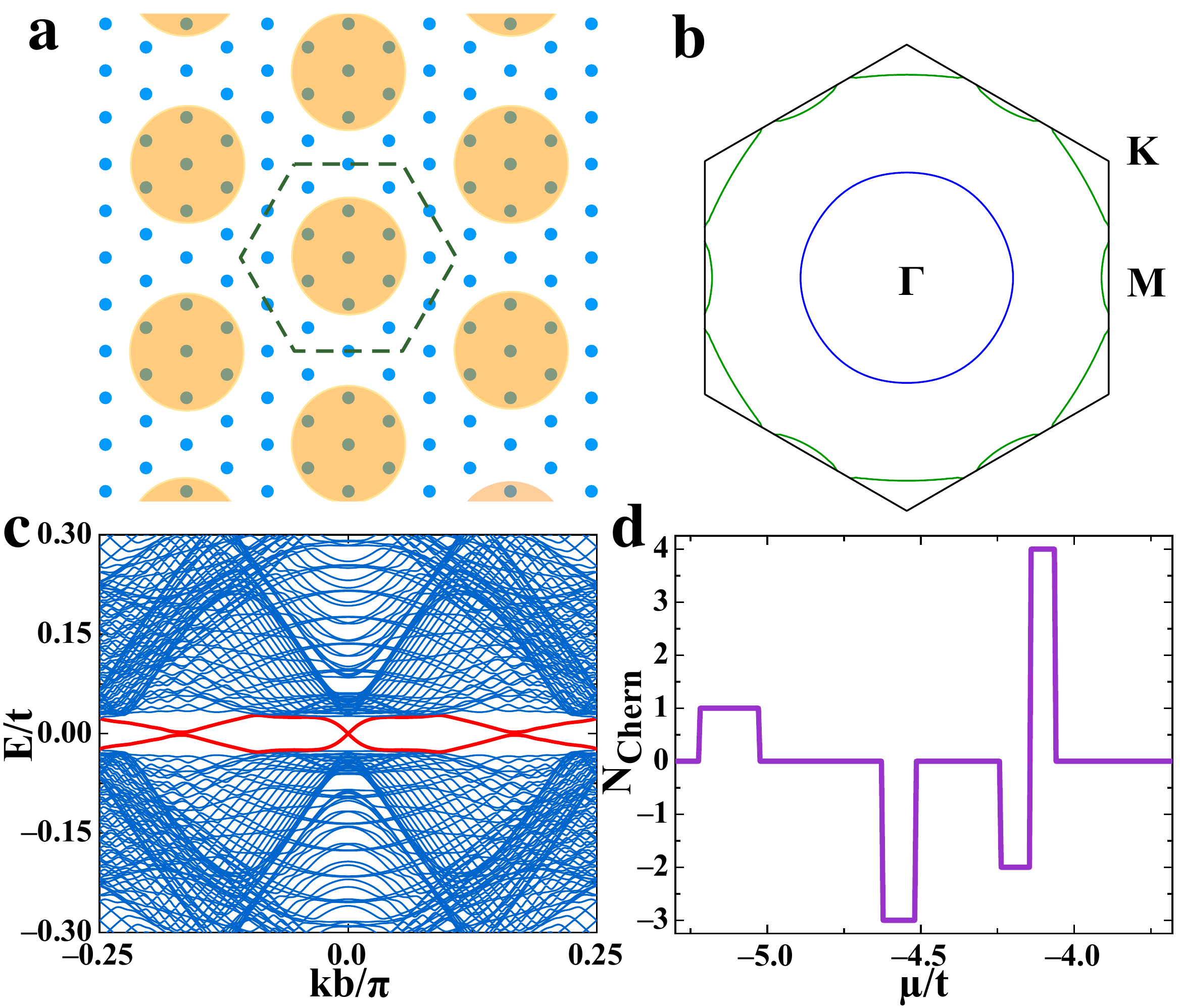}
\caption{ (a) The schematic illustration of organic molecules, denoted by the disks, on Au (111) surface. The local potential energy of the Au under the molecules are modified. A hexagonal unit cell with 16 Au sites, used in the tight-binding model, is shown.  (b) The hexagon denotes the reduced Brillouin zone due to the superlattice potential. The superlattice is chosen such that the $M$ points touch the Fermi surface of the outer Rashba band. (c) The energy spectrum of a superconductor with Chern number 3. The superconductor is obtained by adding superconductivity and a Zeeman field to a system with reduced Brillouin zone given in (b). (d) The Chern number as a function of chemical potential for a fixed superlattice configuration.}
\label{Fig4}
\end{center}
\end{figure}

Importantly, a Zeeman field can close the bulk gap at the time-reversal invariant points X and Y and reopen it again as illustrated in Fig.3c. After the gap reopening, the Chern number of the outer subband is changed by two, from 1 to -1. As a result, the total Chern number of the occupied bands is -2. The energy spectrum of the system with open boundary condition in the y-direction and periodic boundary condition in the x-direction is shown in Fig.3e. It is evident that in-gap Majorana edge states appear in the energy spectrum. The tight-binding model used is described in the Method section.

Interestingly, if the superlattice potential is rectangular such that only the energy gap at X point in Fig.3b is closed and reopened by the Zeeman field, the Chern number of the outer band is zero. As a result, the total Chern number of the system is -1 instead.

\emph{Triangular lattice}--- One of the great advantages of the superlattice scheme is that many different lattice structures can be constructed on the metal surfaces. An interesting possibility is to construct a triangular lattice on top of Au(111) surface as depicted in Fig.4a. If the superlattice spacing is approximately $\pi/k_F$, the new Brillouin zone will have hexagonal shape and the three M points will touch the outer Rashba bands of the Au(111) surface states as shown in Fig.4b. Adding an s-wave pairing potential will gap out the Fermi surfaces and the total Chern number of the occupied bands is zero. However, increasing the Zeeman field can close and re-open the gap at the three M points such that the Chern number of the outer band is changed by 3 from 1 to -2. As a result, the total Chern number of the occupied bands is -3. This results in three chiral edge states at a zig-zag edge of the sample as shown in Fig.4c. The Chern number as a function of chemical potential for a fixed superlattice configuration is shown in Fig.4d.

\section{\bf Discussion}

In the above sections, we discussed how non-magnetic superlattice potentials and uniform magnetic field can be used to engineer topological superconductors in systems with Rashba split bands. In particular, we pointed out that the superlattice potential on Au(111) surface can be induced by non-magnetic coronene molecules. However, a strong external uniform magnetic field is not necessary.  Indeed, the non-magnetic molecules can be replaced by organic molecules with magnetic centers. For example, it has been shown that Co-phthalocyanine and Fe-porphyrin adsorbed on Au(111) surface carries magnetic moments and exhibit Kondo resonances [\onlinecite{Kondo1, Kondo2, Kondo3}]. As a result, the moments can be aligned by a weak external magnetic field. These ferromagnetic superlattice potential can result in topological superconductors similar to uniform magnetic field with non-magnetic superlattices. 

It is also important to note that the superlattice method proposed in this work is very general. The Au(111) surface can be replaced by other 2D systems such as Bi (111) [\onlinecite{Bi}] and Sb (111) [\onlinecite{Sb, David}] surfaces which also support Rashba bands. 

Another possible way to realize topological superconductors is to place magnetic molecules on superconducting Pb thin films. It has been shown that Pb thin films on Si substrate possess Rashba split bands [\onlinecite{Pb1}] and become superconducting at low temperatures [\onlinecite{Pb2, Pb3}]. Importantly, it has been shown that magnetic metalorganic molecules on Pb surfaces can induce Yu-Shiba states [\onlinecite{Pb4}]. Therefore, we believe that superlattice potentials on Pb surfaces induced by magnetic molecules can be used to realize topological superconductors without fine-tuning the chemical potential as demonstrated above.

\section{\bf Method} 

\emph{Calculation of $N_{BDI}$}--- To calculate the 1D topological invariant of $H_{1DTB}$, we note that in the basis which diagonalises $C$, $H_{1DTB}$ can be block diagonalised as:
\begin{equation}
H_{1DTB}(k) = \left(
\begin{array}{cc}
 0 & A(k) \\
A^{T}(-k) & 0
\end{array} \right).
\end{equation}
With the matrix $A(k)$, one can define the phase $\theta$,
\begin{equation}
z(k)=e^{i \theta(k)}=\text{Det}[A(k)]/|\text{Det}[A(k)]|,
\end{equation}
such that $\theta(k) = n \pi$ at $k=0, \pm \pi$ with integer $n$. The winding number of $\theta(k)$ can be used as the topological invariant which characterizes the Hamiltonian $H_{1DTB}(k)$ [\onlinecite{Tewari, Chris}]. The winding number $N_{BDI}$, which counts the number of Majorana end states at one end of the wire, can be written as
\begin{equation}
N_{BDI}=\frac{-i}{\pi} \int_{k=0}^{k=\pi} \frac{dz(k)}{z(k)}.
\end{equation}
$N_{BDI}$ as a function of chemical potential is shown in Fig.2e. When comparing Fig.2b to Fig.2e, it is evident that the system is topological when the chemical potential intercepts a single subband.

\emph{2D tight-binding models} -- The Rashba bands of the  Au(111) surface with the superlattice potential is described by a tight-binding model. The lattice structure of Au (111) surface is triangular as depicted in Fig.4a. To include the effect of the superlattice, we choose a hexagonal unit cell with 16 sites and the on-site chemical potential of the 7 sites inside the muffin-tin circles are modified. The tight-binding model can be written as:

\begin{equation}
\begin{aligned}
H_{2DTB} & =  -t\sum_{\left \langle i,j \right \rangle}\psi_{i}^{\dagger}\psi_{j}-\mu\sum_{i}\psi_i^{\dagger}\psi_i +u\left(i\right)\sum_{i}\psi_i^{\dagger}\psi_i \\
&  -i\alpha_R\sum_{\left \langle i,j \right \rangle,\alpha,\beta}\psi_{i,\alpha}^{\dagger}\left(\mathbf{e}_{i,j}\times \mathbf{\sigma}\right)\cdot \mathbf{e}_z\psi_{j,\beta}\\
& -V_z\sum_{i}\left(\psi_{i\uparrow}^{\dagger}\psi_{i\uparrow}-\psi_{i\downarrow}^{\dagger}\psi_{i\downarrow}\right)  -\Delta\sum_{i}\psi_{i\uparrow}^{\dagger}\psi_{i\downarrow}^{\dagger}+h.c.
\end{aligned}.
\end{equation}

Here, $\mathbf{e}_{ij}$ is a unit vector connecting sites $i$ and $j$, $\mathbf{e}_z$ is a unit vector in the $z$ direction, $u\left(i\right)$ is nonzero for sites inside the muffin-tin circle as depicted in Fig.4a.

In the square lattice case, for simplicity, we have chosen a square lattice tight-binding model to simulate the Rashba bands. A unit cell of 16 sites is used and the superlattice is simulated by changing the local chemical potential of some of the sites in the unit cell, similar to the triangular case. The square lattice model can be justified by the fact that the Fermi surfaces of the Rashba bands of Au (111) surface are circular and far from the original Brillouin zone boundary.  The same circular Fermi surface can be reproduced by square lattice models as long as the Fermi surface is far away from the Brillouin zone boundary.

\section{Acknowledgement} 
We acknowledge the support of HKRGC through HKUST3/CRF/13G and GRF Grants 602813, 605512, 16303014, 603611. We thank Steve Louie, Masatoshi Sato and Yukio Tanaka for discussions.
\emph{Note Added}: At the finishing stage of this work, we noted that there are proposals using 2D arrays of Yu-Shiba states to engineer topological superconductors [\onlinecite{Yu1, Yu2}], even though the importance of band folding was not emphasised in these works.

\end{document}